\documentclass[preprint,nofootinbib,nobibnotes,groupaddress,preprintnumbers,aps,prd]{revtex4}
\usepackage{bm,epsfig,mathrsfs,amsmath,amssymb,graphicx}
\usepackage[toc,title,titletoc,header]{appendix}
\usepackage{graphicx}
\usepackage{dcolumn}
\usepackage{bm}
\usepackage{array}
\newcommand{\PreserveBackslash}[1]{\let\temp=\\#1\let\\=\temp}
\newcolumntype{C}[1]{>{\PreserveBackslash\centering}p{#1}}
\newcolumntype{R}[1]{>{\PreserveBackslash\raggedleft}p{#1}}
\newcolumntype{L}[1]{>{\PreserveBackslash\raggedright}p{#1}}

\def\appendix{}
\def\dbar{{\mathchar'26\mkern-12mu d}}

\usepackage{CJK}
\begin{document}
\begin{CJK*}{GB}{gbsn}
\title{ Efficiency at maximum power of a quantum Otto engine: Both within finite-time and irreversible thermodynamics }
\author{Feilong Wu$^1$}

\author{Jizhou He$^1$} \author{Yongli Ma$^2$} \author{Jianhui Wang$^{1,2}$}\email{wangjianhui@ncu.edu.cn}
\affiliation{ $^1\,$ Department of Physics, Nanchang University,
Nanchang 330031, China}
 \affiliation{ $^2\,$ State Key Laboratory
of Surface Physics and Department of Physics, Fudan University,
Shanghai 200433, China}

\begin{abstract}
We consider the efficiency at maximum power of  a quantum Otto
engine, which uses a spin or a harmonic system as its working
substance and works between two heat reservoirs at constant
temperatures $T_h$ and $T_c$ $ (<T_h)$.  Although the spin-$1/2$
system behaves quite differently from the harmonic system in that
they obey two typical quantum statistics, the efficiencies at
maximum power based on these two different kinds of quantum systems
are bounded from the upper side by the  same expression of the
efficiency at maximum power:
 $\eta_{mp}\leq\eta_+\equiv \eta_C^2/[\eta_C-(1-\eta_C)\ln(1-\eta_C)]$,
 with $\eta_C=1-T_c/T_h$
the Carnot efficiency, which displays the same universality of the
CA efficiency $\eta_{CA}=1-\sqrt{1-\eta_C}$ at small relative
temperature difference. Within context of irreversible
thermodynamics, we calculate the Onsager coefficients and, we show
that the value of $\eta_{CA}$ is indeed the upper bound of EMP for
the Otto engines working in the linear-response regime.

 PACS number(s):
05.70.-a,~03.65.-w
\end{abstract}

\maketitle
\date{\today}

\section{introduction}
Heat engines proceeding in finite time are optimized for powers and
efficiencies within the framework of finite-time thermodynamics,
which was initiated by the seminal paper  of Curzon and Ahlborn
\cite{Cur75}. Under the  assumptions that heat flow obeys the linear
Fourier law and that irreversibility only arises from the heat flow,
Curzon and Ahlborn  considered a Carnot-like heat engine model
working between a hot and a cold reservoir at constant temperatures
$T_h$ and $T_c (<T_h)$,  and they found the efficiency at maximum
power (EMP) to be $\eta_{CA}=1-\sqrt{T_c/T_h}=1-\sqrt{1-\eta_C}$
with $\eta_C=1-T_c/T_h$  the Carnot efficiency. Since then,
subsequent various theoretical papers discussed the bounds and
possible university of the EMP \cite{Esp10, Sch08, Gav10, Sei11,
Gom06, Tu12, Shen14, Guo13, All13}, and some of these studies indeed
disclosed some sort of university of the CA efficiency\cite{Tuz08,
Gom06, All08, Wan13, Jim07, Sch08, Esp10}.

Quantum heat engines supply good model systems to disclose the
emergence of basic thermodynamic description at the quantum
mechanical level, and reveal the relation between the quantum
classical and quantum thermodynamic systems. A large number of
publications  (see, for a review, Refs. \cite{Kos13, Kos14}) have
been devoted to the research into the models of quantum heat engines
proceeding finite time. Among most of these studies, finite-time
thermodynamics as a very useful tool was used to optimize the heat
engines, like the Carnot engine \cite{Wu06,1203}, the Otto engine
\cite{Hug12, Wan13, Rez06, Jian12, Fel00}, and the Brayton engine
\cite{ Huang13, Aba12}, etc. An Otto cycle is reciprocating and
partitioned into four branches, two adiabats, where no heat
exchanges between the working substance and its environment, and two
isochores which are heat transfer processes. Three of the authors
\cite{Wan13}  of the present work optimized a quantum Otto engine
(QOE) model, which uses a two-level atomic system as its working
substance and works between two heat reservoirs at constant
temperatures $T_c$ and $T_h$, and found that the EMP $\eta_{mp}$ is
bounded from the upper side by a function of the Carnot efficiency
$\eta_C$,
\begin{equation}
 \eta_{mp}\leq\eta_{+}\equiv\frac{
\eta_C^2}{[\eta_C-(1-\eta_C)\ln(1-\eta_C)]}=\frac{\eta_C}{2}+\frac{\eta_C^2}{8}+\frac{7\eta_C^3}{96}+O(\eta_C^4),
\label{etcc}
\end{equation}
which was also derived previously in  a steady-state engine model
based on a mesoscopic \cite{Tuz08} or macroscopic \cite{Van12}
system. It is clear that $\eta_{+}$ in Eq. (\ref{etcc}) and
$\eta_{CA}=\frac{\eta_C}{2}+\frac{\eta_C^2}{8}+\frac{6\eta_C^3}{96}+O(\eta_C^4)$
share  the same universality of the EMP at small relative
temperature difference. It is widely believed that the performance
in finite time of a classical Otto cycle depends sensitively on the
working substance \cite{Guo13}.  Here it does raise a very
interesting question deserving to be studied. Is this result
(\ref{etcc}) still valid for the Otto engine which uses other kind
of quantum systems instead of the two level system? To answer this
question, in this paper we use a spin-$1/2$ or a harmonic system
which obeys one of two typical quantum statistics (Fermi-Dirac or
Bose-Einstein) as the working substance of the Otto engine to
determine the EMP.

The relationship between the irreversible thermodynamics and
finite-time thermodynamics was first discussed in Ref. \cite{Van05}
. In his seminar work, Van den Broeck  addressed using the Onsager
relations the generality of the CA efficiency, and proved that
$\eta_{CA}$ is the upper bound of the EMP for heat engines in the
linear response regime $\Delta T\rightarrow0$, with $\Delta
T=T_h-T_c$. Various cyclic or steady-state models of heat engines or
refrigerators, such as  Brwonian motors \cite{Van04, Van06, Izu10},
electronic transport systems \cite{Rut09}, and a macroscopic Carnot
cycle \cite{Izu09}, etc., have been subsequently investigated, in
some of which the Onsager relations have been calculated explicitly
within the framework of the linear \cite{Izu09} or nonlinear
\cite{Izu12, Izu13} irreversible thermodynamics. However, rarely has
the issue of the EMP and of the Onsager coefficients been discussed
for the QOEs. It is therefore of great interest to consider the QOEs
within the framework of irreversible thermodynamics, which may help
us understand the intrinsic relation between the finite-time and
irreversible thermodynamics.

In the present paper, we employ a spin and a harmonic  system as a
working substance to set up a QOE model, which consists of two
isochores and two adiabats.  Optimizing with respect to power of the
QOE, we find that the upper bound of EMP is
$\eta_{+}=\frac{\eta_C}{2}+\frac{\eta_C^2}{8}+\frac{7\eta_C^3}{96}+O(\eta_C^4)$,
which agree well with $\eta_{CA}$. Within the framework of the
irreversible thermodynamics, we prove that the EMP for the is indeed
bounded from the above $\eta_{CA}$, which becomes achievable as the
model satisfy the tight-coupling condition.

\section{Expectation Hamiltonian of a  spin$-1/2$ or a harmonic oscillator system}
\subsection{A spin$-1/2$ an a harmonic oscillator system}
We first consider a quantum system with a magnetic moment
$\textbf{M}$ placed in a magnetic field $\textbf{B}$ whose direction
is assumed to be constant and
along the positive $z$ axis.
The Hamiltonian of the interaction between the magnetic moment
$\textbf{M}$ of the quantum system and the external magnetic field
$\textbf{B}$ is given by $\hat{{H}}(t)=-\textbf{M}\cdot\textbf{ B
}=2\mu_B \textbf{S}\cdot\textbf{B}=2\mu_B {B}_z(t){S}_z,$ where
$\mu_B$ is the Bohr magnetron, $S$ is a spin angular momentum, and
$\hbar=h/(2\pi)$ with $h$ being the Planck constant. Here and
hereafter we adopt $\hbar=1$.  For simplicity, we define
$\omega(t)=2\mu_B {B}_z(t)$. Since the spin angular momentum and
magnetic moment are in opposite directions, the frequency of the
trap $\omega(t)$ must be positive.  Therefore, the Hamiltonian of a
spin$-1/2$  system coupling with the time-dependent field
$\omega(t)$ can be expressed as
\begin{equation}
\hat{H}=\omega(t)\hat{S_z}. \label{hsz}
\end{equation}
In view of the fact that, the expectation value of the spin angular
momentum $S_z$ is given by $S=\langle
S_z\rangle=-\frac{1}2\tanh\left(\frac{\beta\omega}2\right)$, we can
write the  expectation of the Hamiltonian as,
\begin{equation}
\langle \hat{H}\rangle=\omega
S=-\frac{1}2\omega\tanh\left(\frac{\beta\omega}2\right).
\label{ener}
\end{equation}


Let us consider a single harmonic oscillator with time-dependent
frequency $\omega(t)$. The Hamiltonian of the harmonic oscillator is
described by
\begin{equation}
\hat{H}=\omega (t)\left(\hat{N}+\frac{1}2\right)=\omega (t)
\left(\hat{a}^\dag\hat{a}+\frac{1}2\right), \label{ener}
\end{equation}
where $\hat{N}$ is the number operator, and $\hat{a}^\dag, \hat{a}$
are the Bosonic creation and annihilation operators, with
$\hat{N}=\hat{a}^\dag \hat{a}$. The expectation of the Hamiltonian
of the oscillator with inverse temperature $\beta$ is then
 given by
\begin{equation}
\langle \hat{H}\rangle=\omega
{{n}}\equiv{\omega}\left(\bar{n}+\frac{1}2\right)=\frac{1}2\omega\coth\left(\frac{\beta\omega}2\right),
\label{enh}
\end{equation}
where the use of
$\langle\hat{N}\rangle=\bar{n}=\frac{1}{e^{\beta\omega}-1}$ and
${n}\equiv (\bar{n}+\frac{1}2)$ has been made, with $n$ rather than
$\bar{n}$ being used to denote the mean population.

Note that, the expectation  Hamiltonian $\langle \hat{H}\rangle$ of
a system with inverse temperature $\beta$ can be expressed as

\begin{equation}
\langle H\rangle=\omega f(e^{-\beta\omega/2}), \label{hga2}
\end{equation}
where $f$ is the mean population $n$ for the harmonic system or the
mean polarization $S$ for the spin$-1/2$ system.
\subsection{Motion equation of the system Hamiltonian}
The cycle of operation of the QOE is  composed of two adiabats and
two isochores. The quantum dynamics are generated by external fields
during  the two adiabatic processes  and by heat flows from hot and
cold reservoirs in the two isochoric processes. Based on a semigroup
approach, the change in time of an operator $\hat{X}$ during the
adiabatic and the isochoric processes is described by the quantum
master equation \cite{Gev92, Rez06}:
\begin{equation}
\frac{d{\hat{X}}}{d t}=i[\hat{H},~\hat{X}]+\frac{\partial
{\hat{X}}}{\partial t}+\mathcal{L}_D(\hat{X}), \label{xtdx}
\end{equation}
where $\mathcal{L}_D(\hat{X})=\sum_\alpha k_\alpha
\left(\hat{V}_\alpha^\dag\left[\hat{X},
\hat{V}_\alpha^\dag\right]+\left[\hat{V}_\alpha^\dag,\hat{X}
\right]\hat{V}_\alpha\right)$ represents the Liouville dissipative
generator when the system is coupled to a heat reservoir. Here
$\hat{V}_\alpha^\dag$ and $\hat{V}_\alpha$ are operators in the
Hilbert space of the system and are Hermitian conjugates, and
$k_\alpha$ are phenomenological positive coefficients. When
$\hat{X}=\hat{H}$, the internal energy of the system is  of the
expectation value of the Hamiltonian, i.e., $E=\langle
\hat{H}\rangle$. Then substituting $\hat{H}$ into Eq. (\ref{xtdx})
leads to the quantum version of the first law of thermodynamics $dE
= dW + dQ$,
\begin{equation}
\frac{d{{E}}}{d t}=\frac{\dbar{W}}{d t}+\frac{\dbar{Q}}{d
t}={\left\langle\frac{\partial{\hat{H}}}{\partial
t}\right\rangle}+\langle\mathcal{L}_D(\hat{H})\rangle. \label{dote}
\end{equation}
The power and the instantaneous heat flow are identified as, $P =
\frac{\dbar{W}}{d t} =
{\left\langle\frac{\partial{\hat{H}}}{\partial t}\right\rangle}$ and
$\frac{\dbar{Q}}{d t}=\langle\mathcal{L}_D(\hat{H})\rangle$,
respectively.


The operators $\hat{V}^\dag$ and $\hat{V}$, are chosen as the
Bosonic (spin) creation $\hat{a}^\dag$ ($\hat{S}^\dag=\hat{S}_x+i
\hat{S}_y$) and annihilation operators $\hat{a}$
($\hat{S}=\hat{S}_x-i \hat{S}_y$) for the harmonic oscillator
(spin$-1/2$) system. Substituting $\hat{X}
=\omega(\hat{a}^\dag\hat{a}+\frac{1}2)~
 (\hat{X} =\omega\hat{S}_z)$ into Eq. (\ref{xtdx}) and taking the
expectation value leads to the motion of the system Hamiltonian,
\begin{equation}
\frac{d\langle H\rangle}{dt}
 =-\gamma(\langle H\rangle-\langle{H}\rangle^{eq}),
 \label{dn}
\end{equation}
where $\gamma=k_\downarrow-{k_\uparrow}
~(\gamma=k_\downarrow+{k_\uparrow})$ is heat conductivity for the
harmonic (spin) system and
$k_\uparrow/k_\downarrow=e^{-\beta\omega}$ obeys the detailed
balance  ensuring that the system evolves in a specific way to the
correct equilibrium state asymptotically \cite{Fel00}. Here
${\langle H
\rangle}^{eq}=\omega\widetilde{n}^{eq}=\frac{\omega}2\frac{k_\downarrow+k_\uparrow}{k_\downarrow-{k_\uparrow}}$
(or $\langle
H\rangle^{eq}=\omega{S}^{eq}=\frac{\omega}2\frac{k_\downarrow-k_\uparrow}{k_\downarrow+{k_\uparrow}}$)
is the asymptotic value of $\langle H\rangle$.  This asymptotic
population must correspond to the value at thermal equilibrium:
$\widetilde{n}=\frac{1}2\coth(\beta\omega)$ [or $S=-\frac{1}2
\tanh(\beta\omega)$].

\section{quantum Otto cycle}



It follows, using Eq. (\ref{hga2}) and (\ref{dote}), that  for a
spin$-1/2$ or for a harmonic system the first law of thermodynamics
can be expressed as
\begin{equation}
d E=\dbar W+\dbar Q=fd\omega+\omega df, \label{dedn}
\end{equation}
where  $\dbar Q=\omega df$ and $\dbar W=f d\omega$.  The energy of
the system can change either by  particle transition from one level
to the other (changing ${f}$) or by varying the energy gap between
the energy levels (changing $\omega$). It is clear that, a
thermodynamic process, during which the ratio $f(\omega, T)$
remains constant, is a quantum adiabatic process. 
Based on quantum adiabatic theorem \cite{Bor28}, a system would
remain in its initial state during an adiabatic process, but  it
must fulfill the condition that the time scale of its state change
must be much larger than that of the dynamical one, $\sim E/\hbar$.
That means, the time required for completing a quantum adiabatic
process should be very large and cannot be negligible. Therefore, we
must consider nonadiabatic dissipation \cite{Wan13, Fel00} (due to
rapid change of the system energy level) and in particular, the time
taken for any quantum adiabatic process.

Because of nonadiabatic dissipation, the heat is  developed and
yields an increase in entropy in an ``adiabatic'' process which
becomes non-isentropic. 
 In what
follows, even if there exists nonadiabatic dissipation in an
``adiabatic'' process, we still use the word ``adiabatic'' to merely
indicates that the working substance, isolated from a heat
reservoir, has no heat exchange with its surroundings.

A irreversible QOE  cycle $1\rightarrow 2\rightarrow 3\rightarrow
4\rightarrow1$ based on a harmonic system is drawn in the $(\omega,
n)$ plane, as shown in Fig. 1. (Similar schematic diagram, which can
be seen in Ref. \cite{Fel00}, is not plotted here for the QOE based
on a spin system). During two isochoric processes $1\rightarrow 2$
and $3\rightarrow 4$, the working system, at constant volume
$\omega_b$ and $\omega_a$, is coupled to a hot and a cold heat
reservoir whose temperatures are $T_h$ and $T_c$, respectively. Let
$f_i$ be the populations or polarizations at the instants $i$ with
$i = 1, 2, 3, 4.$  During the adiabatic process $2 \rightarrow3$
($4\rightarrow 1$), the working substance is decoupled from the hot
(cold) reservoir, and $f$ changes from $f_2$ to $f_3$ ($f_4$ to
$f_1$). The cycle model is operated in the following  four branches.


\begin{figure}[tb]
\includegraphics[width=2.8in]{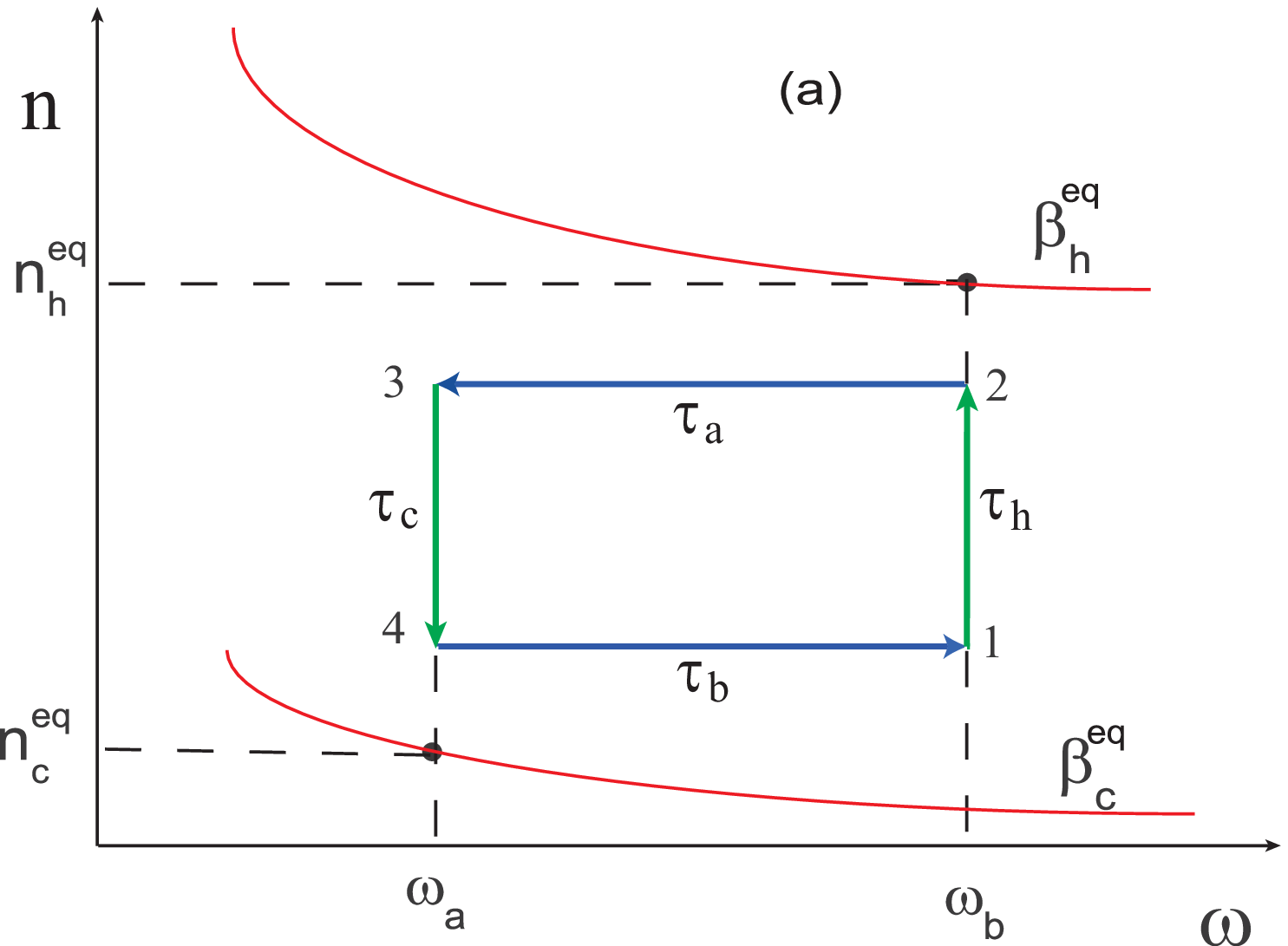}
\includegraphics[width=2.8in]{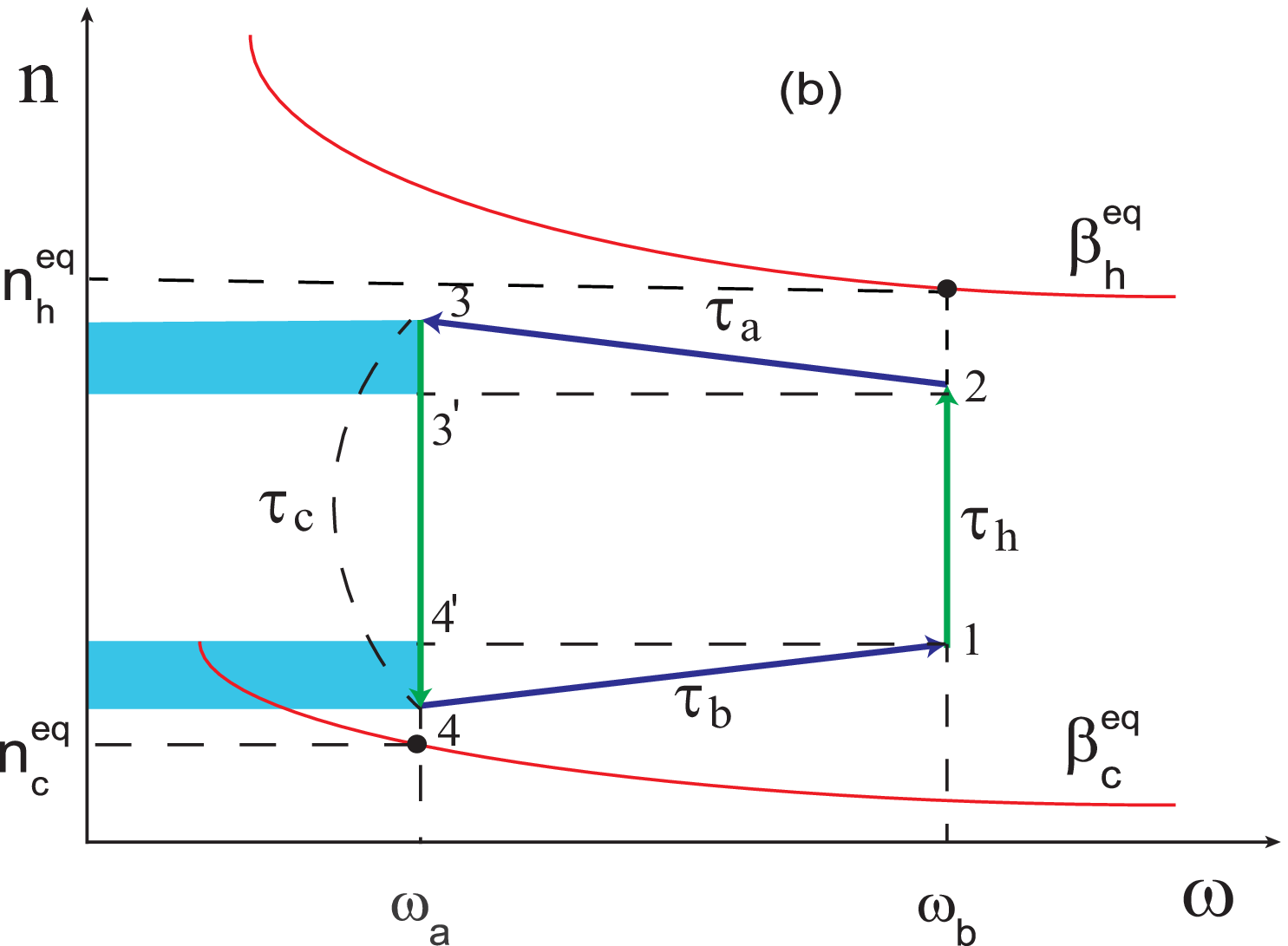}
 \caption{(Color online)  Schematic diagram of a quantum Otto cycle working with a harmonic system in the $(\omega, n)$
 plane. (a): without nonadiabatic dissipation, (b): with nonadiabatic dissipation.
$1\rightarrow 2$ and $3\rightarrow 4$ are two isochoric processes,
while $2\rightarrow 3$ and $4\rightarrow 1$ are two adiabatic
processes.
 ${n}_h^{eq}$ and ${n}_c^{eq}$ are  two ratios of the atomic system
 at thermal equilibrium with two heat reservoirs at  inverse temperatures $\beta_h$ and $\beta_c$.}\label{cyc}
\end{figure}

\emph{1}. Hot isochore 1$\rightarrow$ 2.  The  magnetic field
$\omega$ is kept fixed at constant value of $\omega_b$ and no work
is done. The working subsystem is in contact with  a hot heat
reservoir at inverse temperature $\beta_h$ during a period
$\tau_{h}$, with $\beta_h=1/T_h$.  It follows, using Eq.
(\ref{dedn}), that the instantaneous heat flow becomes
\begin{equation}
\frac{\dbar Q_{12}}{dt}=\omega_b\frac{df(t)}{dt}
=\gamma_h\left[{f}_h^{eq}-f(t)\right]\omega_b, \label{qhft}
\end{equation}
where $\gamma_h$ denotes the heat conductivity between the working
substance and the hot reservoir and  $f_h^{eq}$ is the population
(or polarization) of the harmonic (or spin) system at thermal
equilibrium with the hot reservoir.

In view of the boundary conditons that $f(0) = {f}_1(\omega_b,
\beta_1)$ and ${f}(\infty) = {f}_h^{eq}(\omega_b, \beta_h)$, the
general solution of Eq. (\ref{qhft}) can be readily obtained,
${f}(t)={f}_h^{eq}+({f}_1-f_h^{eq})e^{-\gamma_ht}$, resulting in the
following relation,
\begin{equation}
{f}_2=f_h^{eq}+(f_1-{f}_h^{eq}). \label{n2qx}
\end{equation}
Then the heat absorbed directly from the system in the isochoric
process becomes
\begin{equation}
Q_h\equiv Q_{12}=E_2-E_1=(f_2-{f}_1)\omega_b. \label{qhab}
\end{equation}

\emph{2}. Adiabatic expansion $2\rightarrow 3$. The system is
decoupled from the hot reservoir, changing $\omega$ from $\omega_b$
to $\omega_a$ during time $\tau_a$. Heat caused by the work to
overcome the dissipation  is developed and the population or
polarization is increased from $f_2$ to $f_3$, though there is no
heat exchanged directly between the system and its surroundings. As
in the low-dissipation case \cite{Esp10, Fel00, Wan13}, we assume
that the increase of population or of polarization in an adiabatic
process is inversely proportional to be the time required for
completing this process. Then, we have
\begin{equation}
f_3=f_2+\sigma_a/\tau_a, \label{f3ua}
\end{equation}
where $\sigma_a$ denotes the dissipation coefficient for the
adiabatic expansion. The work done directly during this process,
$W_{23}^0$, can be determined according to
\begin{equation}
W_{23}^0=\int_0^{\tau_a}f
d\omega=(\omega_a-\omega_b)\left(f_2+\frac{\sigma_a}{2\tau_a}\right),
\label{w23}
\end{equation}
while the heat generated on this process in the working substance
becomes
\begin{equation}
W_{23}^{add}=Q_{23}=\int_0^{\tau_a}\omega
df=\frac{\sigma_a(\omega_a+\omega_b)}{2\tau_a},
\end{equation}
which is the additional work to overcome the nonadiabatic
dissipation. That is, the total work done on the adiabatic
compression $2\rightarrow3$, $W_{23}^{add}$, is given by
\begin{equation}
W_{23}=W_{23}^0+W_{23}^{add}=(\omega_a-\omega_b)\left(f_2+\frac{\sigma_a}{2\tau_a}\right)+\frac{\sigma_a(\omega_a+\omega_b)}{2\tau_a}
\end{equation}

{\emph{3}.} Cold isochore $3\rightarrow 4$. The system becomes
coupled to a cold reservoir at inverse temperature $\beta_c$
($>\beta_h$) in a time of $\tau_c$. In a way similar to that for the
step $1\rightarrow2$, the heat current in this process can be given
by
\begin{equation}
\frac{\dbar Q_{34}}{dt}=\omega_a\frac{df(t)}{dt}
=\gamma_c\left[{f}_c^{eq}-f(t)\right]\omega_a, \label{qhv}
\end{equation}
thereby yielding the following relation,
\begin{equation}
{f}_4={f}_c^{eq}+({f}_3-f_c^{eq}) e^{-\gamma_c \tau_c}. \label{n4qy}
\end{equation}
Here $\gamma_c$ is the heat conductivity between the working
substance and the cold reservoir and  $n(t)$ should be restricted by
the boundary constraints: $n(0) = {n}_2(\omega_a, \beta_3)$ and
${n}(\infty) = {n}_c^{eq}(\omega_a, \beta_c)$. The amount of heat
absorbed by the system from the cold reservoir can be directly
calculated as,
\begin{equation}
Q_c\equiv Q_{34}=\left|\int_0^{\tau_b}\omega_ad f \right|=\omega_a\left[(f_2-f_1)+\frac{\sigma_a}{\tau_a}+\frac{\sigma_b}{\tau_b}\right]. \\
\label{qc}
\end{equation}

{\emph{4.}} Adiabatic compression $4\rightarrow 1$. The frequency
$\omega$ is changed from $\omega_b$ to its initial value $\omega_a$
after time $\tau_b$, while $f$ increase from $f_4$ to $f_1$.  The
time required for completing this adiabat is $\tau_b$. As in the
adiabatic expansion $1\rightarrow2$, we assume
\begin{equation}
f_1=f_4+\frac{\sigma_b}{\tau_b},
\end{equation}
with $\sigma_b$ the dissipation coefficient for the process. It
follows, using the computation similar to that for the adiabatic
expansion, that the work done and the heat generated on this adiabat
are,
\begin{equation}
W_{41}^0=\int_0^{\tau_b}fd\omega=(\omega_b-\omega_a)\left(f_1-\frac{\sigma_b}{2\tau_b}\right),
\label{w41}
\end{equation}
\begin{equation}
W_{41}^{add}=Q_{41}=\frac{\sigma_b(\omega_a+\omega_b)}{2\tau_b},
\end{equation}
respectively. Then the total work done on this process is,
\begin{equation}
W_{41}=W_{41}^0+W_{41}^{add}=(\omega_b-\omega_a)\left(f_1-\frac{\sigma_b}{2\tau_b}\right)+\frac{\sigma_b(\omega_a+\omega_b)}{2\tau_b}
\end{equation}

 Repeatedly performing the above sequence of consecutive steps leads to the result that,
 some of heat systematically extracted
from the hot reservoir  is released to the cold reservoir, while the
rest of the heat is delivered as work. After a single cycle, the
total energy of the system as a  state function remains unchanged,
namely, $\Delta E=Q_h-Q_c+W_{23}+W_{41}=0$. The total work done by
the system per cycle, with $W=-(W_{23}+W_{41})$,  and the efficiency
are, respectively, given by
\begin{equation}
   W_{cylce}=
({f}_2-{f}_1)(\omega_b-\omega_a)-\omega_a \left(
\frac{\sigma_a}{\tau_a}+\frac{\sigma_b}{\tau_b}\right), \label{wcy}
\end{equation}
\begin{equation}
  \eta=\frac{W}{Q_h}=1-\frac{\omega_a}{\omega_b}-\frac{\omega_a}{\omega_b}\frac{\left(
{\sigma_a}/{\tau_a}+{\sigma_b}/{\tau_b}\right)}{(f_2-f_1)}.
\label{etab}
\end{equation}
On the right hand side of Eq. (\ref{wcy}), the first term represents
the total positive work done by the system, while the second term is
the total negative work done  by the system [indicated by the two
blue areas in Fig. \ref{cyc}(b)] to overcome internal friction in
two adiabats. Eq. (\ref{etab}) shows that the efficiency $\eta$
monotonously decreases as the nonadiabatic dissipation coefficient
$\sigma_{a,b}$ increases. For the remainder of the paper, our
analysis mainly focuses on the case that the nonadiabatic
dissipation is very weak and even vanishing, while the time required
for completing the quantum adiabatic process is quite long in order
for the quantum adiabatic condition to be satisfied.

\section{the efficiency at   maximum power output}
Following the same approach as in \cite{Fel00},  we can derive the
following relations by combing
 Eq. (\ref{n4qy}) with Eq. (\ref{n2qx}), $
{f}_2-{f}_1=g(\tau_c,\tau_h)\Delta {f}^{eq}, $
%
where
\begin{equation}
g(\tau_c,\tau_h)=\frac{(e^{\gamma_c\tau_c}-1)(e^{\gamma_h\tau_h}-1)}{e^{\gamma_c\tau_c+\gamma_h\tau_h}-1}
\label{gt}
\end{equation}
and $\Delta{n}^{eq} ={n}_h^{eq}-{n}_c^{eq}$, with $
\Delta{f}^{eq}=f(e^{-\beta_c\omega_a})-f(e^{-\beta_h\omega_b}).$

Considering $\tau_{cycle}=\tau_c+\tau_h+\tau_{adi}$, with
$\tau_{adi}\equiv\tau_a+\tau_b$ the total time required for
completing the two adiabatic processes, and using Eq. (\ref{wcy}),
we can derive the power output as,
\begin{equation}
P=\frac{W}{\tau_{cycle}}=\frac{1}{\tau_{cycle}}\left[(\omega_b-\omega_a)\Delta{f}^{eq}g(\tau_c,\tau_h)-\omega_a
\left(
\frac{\sigma_a}{\tau_a}+\frac{\sigma_b}{\tau_b}\right)\right].
\label{pow}
\end{equation}
We find that, from Eq. (\ref{pow}), the positive work condition is
\begin{equation}
\Delta
f^{eq}>\frac{\omega_a}{(\omega_b-\omega_a)}\frac{(\sigma_a/\tau_a+\sigma_b/\tau_b)}{g(\tau_c,\tau_h)},
\label{wfaa}
\end{equation}
which must be satisfied in order that our engine model can produce
positive work. 
In the ideal case when the adiabatic process is isentropic and thus
$\sigma_a=\sigma_b=0$, the power output in Eq. (\ref{pow}) and the
positive work condition in Eq. (\ref{wfaa}) then simplify to
\begin{equation}
P=\frac{g(\tau_c,\tau_h)}{\tau_{cycle}}
(\omega_b-\omega_a)\Delta{f}^{eq},\label{pfeq}
\end{equation}
\begin{equation}
\frac{\omega_b}{\omega_a}<\frac{\beta_c}{\beta_h}=\frac{T_h}{T_c},
\label{frtc}
\end{equation}
respectively. Note that, the positive condition (\ref{frtc})
confirms the Carnot's theorem.

It can be seen from Eq. (\ref{etab}) that,   the efficiency
increases monotonously  with decrease in the dissipation
coefficients $\sigma_{a,b}$ and approaches the upper bound,
$\eta_{+}$, when $\sigma_{a,b}$ are vanishing. Now let us consider
the upper bound of the EMP, which is obtained in the heat engine
with two isentropic processes,  within the assumption that the time
allocations to the two isochores ($\tau_c$ and $\tau_h$) and to the
adiabats $\tau_{adi}$ are given. Based on Eq. (\ref{pfeq}),
optimizing power output becomes equivalent to optimizing two  values
of external fields $\omega_a$ and $\omega_b$. In the Appendix, we
show that, setting $\partial{P}/\partial{\omega_a}=0$ and
$\partial{P}/\partial{\omega_b}=0$,  the EMP can be approximated
analytically by,
\begin{equation}
\eta_{mp}=\frac{\eta_C^2}{\eta_C-(1 - \eta_C )\ln({1-\eta_C})},
\label{etamp}
\end{equation}
whether for a spin-$1/2$ or for a harmonic system. 
This expression of EMP, as one main result of the present paper, was
previously obtained for the heat engine based on a two-level atomic
system \cite{Wan13}, Feynman's ratchet \cite{Tuz08}, or the
classical transport \cite{Van12}. We have proved  in Appendix that
the EMP given by Eq. (\ref{etamp}) holds well in the region of all
finite temperatures, neither restricted to the classical limit when
the temperatures high enough nor to the linear-response regime when
$\Delta(1/\beta)\rightarrow0$ with
$\Delta(1/\beta)=1/\beta_h-1/\beta_c$. It is interesting to note
that, in contrast to the classical Otto engine where the EMP is
dependent on the working substance \cite{Guo13},  the QOEs based on
a spin or a harmonic system have the same upper bound of the EMP,
which is attainable as nonadiabatic dissipation is vanishing.

Expanding $\eta_{+}$ up to the third term of $\eta_C$ gives rise to
$\eta_{mp}=\eta_C/2+\eta_C^2/8+7\eta_C^3/96+O(\eta_C^4)$, which is
in nice agreement with  the expansion of the CA efficiency
$\eta_{CA}$, with $\eta_{CA}=\eta_C/2+\eta_C^2/8+16\eta_C^3
/96+O(\eta_C^4)$. The values of EMP $\eta_{+}$ derived here are very
close to those of the $CA$ efficiency $\eta_{CA}$, particularly at
small relative difference temperatures they have the same
universality, ${\eta_C}/2+{\eta_C^2}/8$.

\section{Irreversible thermodynamics}
We consider the Onsager relations and the EMP by mapping our model
into  a general linear irreversible heat engine, when the model
proceeds in the linear-response regime. We assume that the heat
engine is working in the linear-response regime where the
temperature difference $\Delta T=T_h-T_C$ is very small. The work is
performed under an external force $F$ and it is determined by $W=F
x$, where $x$ is the thermodynamically conjugate variable of $F$. In
the linear-response regime with $\Delta T\rightarrow 0$,  a
thermodynamic force $X_1=F/T_c\simeq F/T,$ where
$T\equiv(T_c+T_h)/2$ and its conjugate flux $J_1=\dot{x}$. We also
define the inverse temperature difference $1/T_c-1/T_h\simeq\Delta
T/T^2$ as another thermodynamic force $X_2$ and the heat flux
$\dot{Q}_h$ as its conjugate flux $J_2$.

The Onsager relations are used to describe these fluxes and forces
as:
\begin{equation}
J_1= L_{11}X_1+L_{12}X_2, \label{j1x2}
\end{equation}
\begin{equation}
J_2=L_{21}X_1+L_{22}X_2, \label{j2x2}
\end{equation}
where $L_{ij}$'s  are the Onsager coefficients with the symmetry
relation $L_{12} = L_{21}$. Since the entropy variation of working
substance which comes back to its original state is vanishing for
our engine model after a whole cycle, the entropy production rate
$\dot{\sigma}$ can be expressed as $
\dot{\sigma}=-\frac{\dot{Q}_h}{T_h}+\frac{\dot{Q}_c}{T_c}=-\frac{\dot{W}}{T_h}+{\dot{Q}_c}({\frac{1}{T_c}
-\frac{1}{T_h}}), $ where  the dot denotes a quantity  quantity
divided by the cycle
 period $\tau_{cycle}$. In the linear response regime where $\Delta T\rightarrow 0$, $\dot{\sigma}$ can be approximated
by
\begin{equation}
\dot{\sigma}\simeq-\frac{{W}}{T}\frac{1}{\tau_{cycle}}+{\dot{Q}_c}\frac{\Delta
T}{T^2},
\end{equation}
where  the higher terms like $O(\Delta T \dot{W} )$ and $O(\Delta
T^3 \dot{Q}_c)$ have been neglected.  Considering the decomposition
$\dot{\sigma} =J_1X_1+J_2 X_2$, we can define the thermodynamic
\cite{Izu09, Izu10, Shen14}. force as
\begin{equation}
X_1 = -\dot{W}/{T}, X_2 = {\Delta T}/{T^2}, \label{x1t2}
\end{equation}
and their conjugate thermodynamic forces
\begin{equation}
J_1 =  1/{\tau_{cycle}}, J_2=\dot{Q}_c. \label{j2qh}
\end{equation}

Considering the Carnot's theorem, we have
$\eta=1-{\omega_a}/{\omega_b}\leq\eta_C=1-{T_c}/{T_h}$, and in the
linear response regime
 \begin{equation}
 \Delta
\omega/\omega\leq {\Delta T}/{T}. \label{dett}
\end{equation}
Here and hereafter we use $\Delta\omega=\omega_b-\omega_a$ with
$\omega\equiv(\omega_a+\omega_b)/2$.  When the QOE works in the
linear response regime $\Delta T\rightarrow 0$ but it can still
produce positive work, even the Carnot efficiency $\eta_C\simeq
{\Delta T}/{T}$ (as the upper bound of the efficiency $\eta$) tends
to be vanishing, implying that we may assume $\Delta
\omega/\omega\simeq \Delta T/T\rightarrow 0$.

We turn to the explicit calculation of the Onsager coefficients
$L_{ij}$'s, adopting an approach similar to ones used in theoretical
models of a Brownian and a macroscopic Carnot cycle \cite{Izu09,
Izu10}.  To determine $L_{11}$, we consider the relation between
$1/{\tau_{cycle}}$ and $X_1$ in the case of $\Delta T \rightarrow 0$
as well as $\Delta \omega\rightarrow 0$. For simplicity, we assume
$\sigma_a\equiv\sigma_b\equiv\sigma/4$,
$\tau_a\equiv\tau_b\equiv\tau_{cycle}/\alpha$ with $\alpha>1$ in the
following. From Eq. (\ref{qc}), then the amount of heat released to
the cold reservoir becomes
\begin{equation}
Q_c=\omega_a\Delta f^{eq}
g-\frac{\omega_a\sigma\alpha}{\tau_{cycle}}. \label{qc2}
\end{equation}
Since $\Delta T\rightarrow 0$, from Eq. (\ref{pow}) we can write
using $\Delta \omega\rightarrow0$ the work $W$ as
\begin{equation}
W=\Delta \omega \Delta f^{eq}
g-\frac{\omega_a\sigma\alpha}{\tau_{cycle}}. \label{wdai}
\end{equation}

 Setting $\Delta\omega=\Delta T=0$ in Eq. (\ref{wdai}), we have by
using the approximation $\omega_a \simeq\omega_b \simeq\omega$,
\begin{equation}
\frac{1}{\tau_{cycle}}=\frac{T}{\omega\alpha\sigma}\frac{-W}T,
\label{frwt}
\end{equation}
which, together with Eqs. (\ref{j1x2}) and (\ref{x1t2}), gives rise
to
\begin{equation}
L_{11}=\frac{T}{\omega\sigma\alpha}. \label{l11}
\end{equation}

Likewise $\dot{Q}_c$ at $\Delta T = 0$ can  be expressed  by using
Eqs. (\ref{qc2}) and (\ref{frwt})as
\begin{equation}
\dot{Q}_c=\frac{T g{\Delta
f^{eq}}}{\alpha\sigma}\frac{-W}T-\frac{\omega\sigma\alpha}{\tau_{cycle}^2}.
\label{qhab}
\end{equation}
Since the second term in the above equation is  $O(W^2)$ quantity
from Eq. (\ref{frwt}), $\dot{Q}_c$ with $\Delta T=0$ can be
evaluated up to the linear order of $W$,
\begin{equation}
\dot{Q}_c=\frac{ T g{\Delta f^{eq}}}{\alpha\sigma}\frac{-W}T.
\label{qhwt}
\end{equation}
From Eqs. (\ref{j2x2}) and (\ref{x1t2}), the coefficient $L_{21}$ is
determined according to
\begin{equation}
L_{21}=\frac{T g{\Delta f^{eq}}}{\alpha\sigma}. \label{l21}
\end{equation}
Here $g=g(\tau_c,\tau_h)$ defined above Eq. (\ref{gt}) is a function
of the time $\tau_c$ and $\tau_h$, and it is thus a function of the
cycle time $\tau_{cycle}$(=$1/J_1)$. However, the value of parameter
$g$, situated between $0\leq g\leq1$, is dimensionless and it can
then be casted into the expressions of the Onsager coefficients.

  In the linear-response
regime when $\Delta T\rightarrow 0$, we can assume from Eq.
(\ref{dett}) that
 $\Delta \omega/\omega\simeq \Delta T/T\rightarrow 0$; therefore $W$
 in Eq. (\ref{wdai}) is approximately
\begin{equation}
W= {T\omega\Delta f^{eq} g}\frac{\Delta
T}{T^2}-\frac{\omega\sigma\alpha}{\tau_{cycle}}. \label{wdau2}
\end{equation}
When setting $W=0$, we can obtain from Eq. (\ref{wdau2}),
\begin{equation}
\frac{1}{\tau_{cycle}}=\frac{T g{\Delta
f^{eq}}}{\sigma\alpha}\frac{\Delta T}{T^2}. \label{frt2}
\end{equation}
Substitution of $W=0$ into Eq. (\ref{wdai}) leads to
\begin{equation}
L_{12}=\frac{T g{\Delta f^{eq}}}{\sigma\alpha}. \label{l12}
\end{equation}
From Eqs. (\ref{l21}) and (\ref{l12}), we see that the Onsager
symmetry relation $L_{21} = L_{12}$ is confirmed as expected. In the
case of $W=0$, $\dot{Q}_c$ can be derived from Eqs. (\ref{qc2}) and
(\ref{frt2}) as
\begin{equation}
\dot{Q}_c=\frac{\omega T(\Delta
f^{eq})^2g^2}{\sigma\alpha}\frac{\Delta
T}{T^2}-\frac{\omega\sigma\alpha}{\tau_{cycle}^2}.
\end{equation}
Since here the second term is $O(\Delta T^2)$ quantity, we can
neglect this term and then obtain
\begin{equation}
L_{22}=\frac{\omega T(\Delta f^{eq})^2g^2}{\sigma\alpha}.
\label{l22}
\end{equation}

 As expected, these
Onsager coefficients derived in our model satisfy the constraints
$L_{11} \geq 0, L_{22} \geq 0$ and $L_{11}L_{22} -L_{12}L_{21} \geq
0$, which originates from the positivity of the entropy production
rate $\dot{\sigma}$.

Now consider EMP for our linear irreversible heat engine,  following
the approach first proposed in \cite{Van05}. With consideration of
Eqs. (\ref{x1t2}) and (\ref{j2qh}), the power and the efficiency can
be expressed as $ P=\dot{W}=-J_1 X_1 T$, and
$\eta={\dot{Q}_h}/{\dot{W}}=-{J_2}/{(J_1 X_1 T)}$, respectively. It
then follows, using the condition ${\partial P}/{\partial X_1}=0$,
that the EMP takes the form as  $\eta^*=\frac{\Delta
T}{2T}\frac{q^2}{1-q^2}$, where $q=L_{12}/\sqrt{L_{11}L_{22}}$ as
the coupling strength parameter has been used.  These Onsager
coefficients given by Eqs. (\ref{l11}), (\ref{l21}), (\ref{l12}),
and (\ref{l22}) show that here the linear irreversible heat engine
satisfies the tight-coupling condition $|q|=1$. In such case, the
EMP becomes \cite{Izu10}
\begin{equation}
\eta_{mp}=\frac{\Delta T}{2T}=\eta_{CA}+O(\Delta T^2). \label{eta2}
\end{equation}
It is also the upper bound of EMP since the coupling strength
parameter satisfies the relation $|q|\leq 1$ which is equivalent to
the condition that $L_{11} \geq 0, L_{22} \geq 0$ and $L_{11}L_{22}
-L_{12}L_{21} \geq 0$.

\section{Conclusions}

We have  employed both finite-time and irreversible thermodynamics
to consider the EMP for a QOE, in which the working substance is
composed of a spin-$1/2$ and a harmonic system. From a view point of
finite-time thermodynamics, we showed that the EMP, wether for the
spin or harmonic system, is bounded from above the same value
$\eta_{+}$ determined by Eq. (\ref{etcc}) which displays the same
universality as $\eta_{CA}$ at small relative temperature
differences. Within the framework of the linear irreversible
thermodynamics, we proved that $\eta_{CA}$ is the upper bound of the
EMP for the heat engines in linear response regime when the
temperature difference $\Delta T\rightarrow 0$, and we also
calculated the Onsager coefficients for the irreversible QOEs.

 \textbf{Acknowledgements}

This work is supported by the National Natural Science Foundation of
China under Grants No. 11265010, No. 11375045, and No. 11365015; the
State Key Programs of China under Grant No. 2012CB921604; and the
Jiangxi Provincial Natural Science Foundation under Grant No.
20132BAB212009, China.

\begin{appendices}
\numberwithin{equation}{section}
\section{Analytical expression of EMP for a QOE working with a harmonic or a spin-$1/2$ system} \label{pro}
\subsection{For a harmonic system}

In the case when the working substance is a harmonic system, the
power output becomes
\begin{equation}
P=\frac{W}{\tau_{cycle}}=\frac{1}2G(\tau_c,\tau_h)
(\omega_b-\omega_a)[\coth{(\beta_h\omega_b)}-\coth{(\beta_c\omega_a)}].
\label{pfeq}
\end{equation}
Then, the extremal conditions of ${\partial P}/{\partial \omega_a}
=0$ and ${\partial P}/{\partial \omega_b}=0 $ lead to
\begin{equation}
(\omega_b-\omega_a) \beta_c=\left[\coth \left(\frac{\omega_a
\beta_c}2\right) - \coth
\left(\frac{\omega_b\beta_h}2\right)\right]\sinh^2
\left(\frac{\omega_a\beta_c}2 \right), \label{dom1}
\end{equation}
\begin{equation}
(\omega_b-\omega_a) \beta_h=\left[\coth \left(\frac{\omega_a
\beta_c}2\right) - \coth
\left(\frac{\omega_b\beta_h}2\right)\right]\sinh^2
\left(\frac{\omega_b\beta_h}2 \right).\label{dom2}
\end{equation}
Dividing directly both sides of Eq. (\ref{dom1}) by Eq. (\ref{dom2})
and defining $r=\sqrt{\beta_c/\beta_h}$, we have
\begin{equation}
\frac{1}{r}= \frac{{1/x_h} -x_h}{{1/x_c} -x_c}, \label{1r}
\end{equation}
in which $x_h\equiv e^{-\frac{\omega_b \beta_h}2}$ and $x_c \equiv
e^{-\frac{\omega_a \beta_c}2}$. The physical solution to Eq.
(\ref{1r}) can be obtained,
\begin{equation}
x_h=\frac{\sqrt{ x_c^4+( 4 r^2-2) x_c^2 +1  }+x_c^2-1}{2 r x_c},
\label{xhxc}
\end{equation}
from which we expand $x_h$ up to the sixth order:
\begin{equation}
x_h=r x_c - r(r^2 - 1) x_c^3 + r(r^2-1)(2r^2-1) x_c^5+O(x_c^7).
\label{xhc7}
\end{equation}
From Eq. (\ref{xhxc}), we note that the condition
$x_c^2-(4r^2-2)x_c+1>0$ must be satisfied in order for $x_h$ to be a
real number. This condition, together with the fact that $0<x_c<1$,
leads to $0<x_c\leq x_c^+$, where $x_c^+= - \sqrt{( 2r^2-1)^2-1
}+2r^2-1$ is the upper bound of $x_c$. Here $x_c^+$ is the same as
corresponding one derived from the two-level atomic system [see
Appendix in Ref. \cite{Rui13}]. We can think of two effective facts:
(1) the upper bound of $x_c$ decreases quickly with increasing $r$
and rapidly approaches zero, favoring $x_h\simeq r x_c$ when $r\gg1$
and $x_c^+\rightarrow0$; and (2) if $r$ is approximated equal to
$1$, the expansion coefficients on the right side of Eq.
(\ref{xhc7}) becomes vanishing, favoring $x_h\simeq r x_c$ when
$r\rightarrow1$ and $x_c^+\rightarrow1$.
That is, Eq. (\ref{xhc7}) can be simplified as
\begin{equation}
x_h=r x_c. \label{xhc}
\end{equation}
This approximation is valid, but not restricted to  the
linear-response regime $\Delta{1/\beta}\rightarrow0$ with $
\Delta{(1/\beta)}=1/\beta_h- 1/\beta_c$ (i.e., $r\rightarrow1$) or
to the high-temperature limit  when $\beta \omega\ll 1$ (i.e.,
$x_c\rightarrow1$).

When we multiply both sides of Eqs. (\ref{dom1}) and (\ref{dom2}),
we obtain $ \omega_b-\omega_a=(2 \sinh( \omega_a\beta_c -
\omega_b\beta_h)/\sqrt{\beta_c \beta_h}$, or
\begin{equation}
 2\left(\frac{\ln x_c}{\beta_c}- \frac{\ln x_h}{\beta_h}\right)=\frac{x_h/x_c - x_c/x_h}{\sqrt{\beta_c
  \beta_h}},
\label{xmy}
\end{equation}
where $x_c$ and $x_h$ were defined in Eq. (\ref{1r}). Considering
Eqs. (\ref{xhc}) and (\ref{xmy}), we have
\begin{equation}
\ln{x_c}=\frac{(r^2-1) \sqrt{\beta_c \beta_h} + 2 r \beta_c \ln(r)}{
 2 r (\beta_h - \beta_c)}, \label{logxc}
\end{equation}
and
\begin{equation}
\ln{x_h}=\ln x_c+\ln r. \label{logxh}
\end{equation}
Substituting $r = 1/\sqrt{1-\eta_c}$, with the Carnot efficiency
$\eta_c= 1-\beta_h/\beta_c$, into the expression
$\eta^*=1-\frac{\omega_a}{\omega_b}=1-\frac{\beta_h\ln{x_c}}{\beta_c\ln{x_h}}$,
we then derive the analytical expression of EMP [see Eq.
(\ref{etamp})].
\subsection{For a spin$-1/2$ system}
If the working substance is a spin$-1/2$ system, then the power
output for the heat engine becomes
\begin{equation}
P=\frac{W}{\tau_{cycle}}=\frac{1}2G(\tau_c,\tau_h)
(\omega_b-\omega_a)[\tanh{(\beta_c\omega_a)}-\tanh{(\beta_h\omega_b)}].
\label{pfeq}
\end{equation}
We set  ${\partial P}/{\partial \omega_a} =0$ and ${\partial
P}/{\partial \omega_b}=0 $, obtaining
\begin{equation}
(\omega_b-\omega_a) \beta_c=\left[\tanh\left(\frac{\omega_a
\beta_c}2\right) - \tanh
\left(\frac{\omega_b\beta_h}2\right)\right]\cosh^2
\left(\frac{\omega_a\beta_c}2 \right), \label{dom3}
\end{equation}
\begin{equation}
(\omega_b-\omega_a) \beta_h=\left[\tanh \left(\frac{\omega_a
\beta_c}2\right) - \tanh
\left(\frac{\omega_b\beta_h}2\right)\right]\cosh^2
\left(\frac{\omega_b\beta_h}2 \right).\label{dom4}
\end{equation}
Based on Eqs. (\ref{dom3}) and (\ref{dom4}), we  find,  in the same
way that we derived Eqs. (\ref{xhxc}) and (\ref{logxc}),  and that
for the spin$-1/2$ system, optimal relations among $x_c, x_h$ and
$r$ are also determined by Eqs. (\ref{xhxc}) and (\ref{logxc}), and
(\ref{logxh}). As a consequence, the EMP for a heat engine working
with a spin$-1/2$ system can be approximated by Eq. (\ref{etamp}),
the same as one obtained from  the heat engine based on the harmonic
system.

\begin{equation}
\end{equation}
\end{appendices}

\end{CJK*}
\end{document}